\documentclass[summary]{URSIGASS2020}
\usepackage{bm}
\usepackage{aas_macros}
\usepackage{amsfonts}
\usepackage{amsmath}
\DeclareMathOperator*{\argmin}{argmin}

\title{Novel perspectives gained from new reconstruction algorithms}

\author{Luke Pratley\affref{ref1}, Melanie Johnston-Hollitt\affref{ref2},
  and Jason D. McEwen\affref{ref3}}

\affiliation{%
  \aff{ref1}{Dunlap Institute for Astronomy and Astrophysics,  University of Toronto, ON M5S 3H4, Canada}
  \aff{ref2}{International Centre for Radio Astronomy Research (ICRAR), 
Curtin University,
1 Turner Ave., Technology Park, Bentley, 6102,  WA, Australia}
  \aff{ref3}{Mullard Space Science Laboratory (MSSL), 
University College London (UCL), 
Holmbury St Mary, Surrey RH5 6NT, UK}
}

\begin{document}

\maketitle

\begin{abstract}
Since the 1970s, much of traditional interferometric imaging has been built around variations of the CLEAN algorithm, in both terminology, methodology, and algorithm development. Recent developments in applying new algorithms from convex optimization to interferometry has allowed old concepts to be viewed from a new perspective, ranging from image restoration to the development of computationally distributed algorithms. We present how this has ultimately led the authors to new perspectives in wide-field imaging, allowing for the first full individual non-coplanar corrections applied during imaging over extremely wide-fields of view for the Murchison Widefield Array (MWA) telescope. Furthermore, this same mathematical framework has provided a novel understanding of wide-band polarimetry at low frequencies, where instrumental channel depolarization can be corrected through the new $\delta\lambda^2$-projection algorithm. This is a demonstration that new algorithm development outside of traditional radio astronomy is valuable for the new theoretical and practical perspectives gained. These perspectives are timely with the next generation of radio telescopes coming online.
\end{abstract}

\section{Introduction}

New low frequency telescopes such as the Murchison Widefield Array (MWA; \cite{tingay13}) and the Canadian Hydrogen Intensity Mapping Experiment  (CHIME; \cite{chime}) have ultra-wide fields of view that are sensitive to the entire diffuse radio sky, with built in polarimetric capabilities to study the magnetized and high redshift universe in precision at low frequencies. This has proven extremely valuable for new scientifically interesting discoveries and has motived the construction of the next generation of low frequency radio interferometric telescopes like the SKA1-LOW which have ambitious science goals, such as detecting the Epoch of Reionization (EoR; \cite{koo15}) and to probe cosmic magnetic fields in Galactic and extra-galactic structures \cite{mjh15}. Furthermore, the large data sets and instrumental response from these telescopes provide a imaging challenge that needs to be overcome for these telescopes to meet the expectations in sensitivity and quality of interferometric imaging that is critical to meet their science goals (e.g. \cite{bea19}). This is especially vital at low frequencies where telescopes image over extremely wide fields of view, allowing huge regions of the sky to be imaged in a single snapshot.

In this paper, we present recent developments from convex optimization \cite{LP18} and wide-field imaging \cite{PJM19a} that have provided new perspectives on algorithm development that are valuable for the next generation radio astronomy.
These developments included the ability to reconstruct diffuse models of the radio sky without the need for image restoration, the ability to distribute computation for reconstruction and the correction wide-field non-coplanar effects, the use of radial symmetry and adaptive quadrature to speed up $w$-projection kernel calculation, and the development of the $\delta\lambda^2$-projection algorithm for wide-band rotation measure synthesis that can correct instrumental depolarization at low radio frequencies.

\section{Convex Optimization}

Convex optimization is a broad field where tools are developed to find solutions to convex minimization problems, which has been recently applied to algorithm development in interferometric imaging \cite{LP18,Dabbech18}.  
An objective function (the function to be minimized) $f$ is convex if $\forall \bm{x}_1, \bm{x}_2 \in \mathbb{R}^N$ $\forall t \in [0, 1]$ the following holds \cite{kom15} 
\begin{equation}
  f(t\bm{x}_1+(1-t)\bm{x}_2)\leq t f(\bm{x}_1)+(1-t)f(\bm{x}_2)\, .
\end{equation}
Many common minimization problems are convex, such as least squares minimization, and they have the useful property that every local minimum is a global minimum.

Convex optimization problems and algorithms are used in many areas of signal processing and have gained popularity in data science for their ability to impose sparse priors and solve non-smooth minimization problems through the use of simple operations known as proximity operators \cite{kom15}. They can make use of wavelet representations to efficiently fit signals on multiple scales and to add robustness to additive Gaussian noise. For example, the constrained $\ell_1$-regularization problem reads
\begin{equation}
  \bm{x}^\star =\argmin_{\bm{x}} \|\mathsf{\bm{\Psi}}^\dagger \bm{x}\|_{\ell_1}\, {\rm such\, that}\quad \|\mathsf{\bm{\Phi}}\bm{x} -\bm{y} \|_{\ell_2} \leq \varepsilon\, ,
\end{equation}
where $ \bm{x}^\star\in\mathbb{C}^N$ is the solution, $\|\cdot \|_{\ell_1}$ is the sum of absolute values and $\|\cdot \|_{\ell_2}$ is the euclidean norm, $\mathsf{\bm{\Psi}}^\dagger\in\mathbb{C}^{K\times N}$ is a wavelet transform or dictionary of transforms. $\varepsilon \in \mathbb{R}^+$ is the noise bound on the measurements $\bm{y}\in\mathbb{C}^M$, and $\mathsf{\bm{\Phi}}\in\mathbb{C}^{M\times N}$ is the operator that maps from the model image to model visibilities. The representation for the sparsity operator $\mathsf{\bm{\Psi}}^\dagger $ can be a collection of wavelet basis, which has shown to improve reconstruction of diffuse sources when compared to one representation alone. The $\ell_1$-norm encourages sparsity, it reduces over fitting in the sense that solutions with mostly zero wavelets coefficients are encouraged. Many algorithms exist that can solve this problem such as the alternating direction method of multipliers and the primal dual algorithms (see \cite{kom15} for more details). Furthermore, these algorithms can be distributed \cite{ono16,PJM19a,pra19d}.

\section{Wide-field Imaging}
Recent work with distributed convex optimization algorithms has aided to new perspectives for calculating the wide-field interferometric measurement equation. In radio interferometry the planar separation of antenna pairs, otherwise known as "baselines", lie in the complex Fourier plane known as the $uv$-plane. In practice we need to account for a 3-dimensional distribution of antennas which includes a third component, out of the $uv$-plane, the so-called $w$-component. The $w$-component measures the Fourier mode of the image domain signal along the line of sight, and for wide-fields of view this corresponds to probing the curvature of the celestial sphere, this curvature needs to be accounted for during image reconstruction \cite{Cornwell08,PJM19a}. Algorithms for accounting for this curvature already exist, such as faceting \cite{tas18} and the $w$-projection and $w$-stacking \cite{Cornwell08} family of algorithms. Each wide-field correction algorithm has a different strategy for modeling wide-field effects during image reconstruction. The wide-field measurement equation can be written as 
\begin{equation}
  y(u, v, w) = \int x(l, m){\rm e}^{-2\pi i (lu + vm + w(\sqrt{1 - l^2 -m^2} - 1))}\,{\rm d}l {\rm d}m\,,
\end{equation}
where $y$ is the measured visibility for the baseline $(u, v, w)$, $x(l, m)$ is the sky with coordinates $(l, m)$.

One of the greatest difficulties of wide-field imaging is that each measurement can have a different $w$ term, meaning a different complex exponential needs to be multiplied in the image domain for each Fourier coefficient for accurate imaging. The $w$-projection algorithm uses the convolution theorem to apply the image domain multiplication in the Fourier domain using a convolution kernel. The $w$-stacking algorithm splits the measurements up into stacks and performs an image domain multiplication for the average $w^\prime$ and a Fourier transform for each stack. 

\begin{figure}
\includegraphics[trim=20 100 20 90,width=0.45\textwidth]{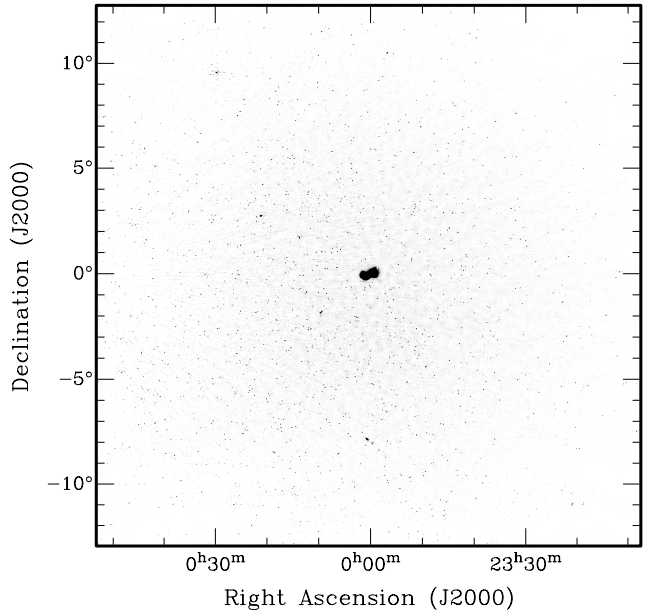}
\caption{A 25 by 25 degree field of view with Fornax A observed using the MWA telescope and imaged using new distributed wide-field image reconstruction \cite{pra19d}. Each of the 100 million measurements has had its individual non-coplanar term accounted for during image reconstruction.}
\label{fig:foxnaxA}
\end{figure}

In our recent work we make new developments to wide-field imaging for each of these components. We show that radial symmetry can reduce the computational cost of the $w$-projection kernel calculation. However, we also show that the kernel calculation can be efficiently and accurately performed using adaptive quadrature, so it needs relatively few samples and is independent of the number of pixels used in the image \cite{PJM19a}.
Furthermore, we use a newly developed $w$-stacking $w$-projection algorithm to image the radio sources Fornax A and Vela where we can apply non-coplanar correction for each visibility over fields of view of approximately 25 by 25 degrees (see Figure \ref{fig:foxnaxA}) \cite{PJM19a, pra19d}. We use MPI (Message Parsing Interface) with the $w$-stacking algorithm to distribute the $w$-stacks over a computing cluster, allowing the average $w$-component to be corrected for each stack in image domain. Then we use the $w$-projection algorithm with radial symmetry to correct for the remaining non-coplanar effect for each visibility in the Fourier domain. In \cite{pra19d} we found that complex conjugation can be used to increase the efficiency of the $w$-stacking algorithm, allowing the first time non-coplanar correction to be performed for over 100 million visibilities in a single observation. This was made possible in conjunction with the new distributed convex optimization algorithms. 

\section{Low Frequency Wide-Band Polarimetry}
Radio polarimetry at low frequencies allows precision measurements of linear polarization that can probe magnetic fields in our Universe more accurately than other wavelengths \cite{Beck13,rise18}. The accuracy of the polarimetry allows precise measurements of wide-band Faraday rotation from a linearly polarized signal, where the linear polarization of light rotates as a function of wavelength squared $\lambda^2$
\begin{equation}
  \chi(\lambda^2) = \chi_0 + \phi\lambda^2\, ,
\end{equation}
where $\chi$ is the polarization angle and $\phi$ is the rotation measure (RM) determined by the integrated magnetic field along the line of sight, specifically
\begin{equation}
  \phi \propto \int n_e(\bm{l})\bm{B}(\bm{l})\cdot {\rm d}\bm{l}\, ,
\end{equation}
where $n_e(\bm{l})$ is the electron density and $\bm{B}(\bm{l})$ is the magnetic field vector \cite{bur66}. There can be different magnetized mediums along the line of sight leading to multiple rates of rotation in the linearly polarized signal. These magnetized mediums can be probed through the use of the rotation measure synthesis measurement equation \cite{bur66,bre05}
\begin{equation}
  P(\lambda^2) = \int P(\phi){\rm e}^{2i \phi \lambda^2} {\rm d}\phi\, ,
\end{equation}
where $P(\lambda^2)$ and $P(\phi)$ is the complex linear polarization signal as a function of $\lambda^2$ and $\phi$ respectively.
The limited number of channels as a function of $\lambda^2$ describes an inverse problem that is mathematically identical to 1-dimensional interferometry.
We describe and show for the first time how at low radio frequencies channel averaging over long wavelengths can result in vector averaging for RM values caused by strong magnetic fields \cite{pratley20}. This creates a decrease in sensitivity to large values of $\phi$ in the new measurement equation for wide-band rotation measure synthesis
\begin{equation}
  P(\lambda^2, \delta \lambda^2) = \int a(\phi, \delta\lambda^2)P(\phi){\rm e}^{2i \phi \lambda^2} {\rm d}\phi\, ,
\end{equation}
where $\delta \lambda^2$ is the channel width that determins the averaging at wavelength $\lambda$, and $a(\phi, \delta\lambda^2)$ describes the channel width dependent depolarization as a function of Faraday depth $\phi$. We show this by drawing the connection between the $w$ and $a$-projection \cite{Cornwell08,Bhat08} algorithms from interferometric imaging and develop a new $\delta\lambda^2$-projection algorithm to correct for the channel averaging depolarization during rotation measure synthesis. We then use the primal dual algorithm to solve for a solution $P(\phi)$ that has been corrected for both missing measurements of $\lambda^2$ and the depolarization due to each channel width $\delta\lambda^2$. This is timely with large RMs becoming increasingly scientifically interesting with strong magnetic fields being poorly understood. Prime examples are from where this will play an extremely important role are newly detected Fast Radio Bursts (FRBs) \cite{aka16} and the strong excess of Faraday rotation in the Sagittarius spiral arm \cite{Shanahan19}.

\section{Conclusions}
Recent convex optimization algorithms have allowed interferometric imaging to be developed in a new perspective. This has allowed novel algorithm development that is designed to operate on high performance computing clusters. Furthermore, the methodology and understanding of wide-field imaging and polarimetry are becoming increasingly scientifically important for the next generation of radio telescopes, and a new perspective is needed if the challenges of imaging and polarimetry are to be met.

\section{Acknowledgements}
The Dunlap Institute is funded through an endowment established by the David Dunlap family and the University of Toronto.
\bibliographystyle{IEEEtran}
\bibliography{refs}

\begin{thebibliography}{10}
\providecommand{\url}[1]{#1}
\csname url@samestyle\endcsname
\providecommand{\newblock}{\relax}
\providecommand{\bibinfo}[2]{#2}
\providecommand{\BIBentrySTDinterwordspacing}{\spaceskip=0pt\relax}
\providecommand{\BIBentryALTinterwordstretchfactor}{4}
\providecommand{\BIBentryALTinterwordspacing}{\spaceskip=\fontdimen2\font plus
\BIBentryALTinterwordstretchfactor\fontdimen3\font minus
  \fontdimen4\font\relax}
\providecommand{\BIBforeignlanguage}[2]{{%
\expandafter\ifx\csname l@#1\endcsname\relax
\typeout{** WARNING: IEEEtran.bst: No hyphenation pattern has been}%
\typeout{** loaded for the language `#1'. Using the pattern for}%
\typeout{** the default language instead.}%
\else
\language=\csname l@#1\endcsname
\fi
#2}}
\providecommand{\BIBdecl}{\relax}
\BIBdecl

\bibitem{tingay13}
S.~J. {Tingay}, R.~{Goeke}, J.~D. {Bowman}, D.~{Emrich}, S.~M. {Ord}, D.~A.
  {Mitchell}, M.~F. {Morales}, T.~{Booler}, B.~{Crosse}, R.~B. {Wayth}, C.~J.
  {Lonsdale}, S.~{Tremblay}, D.~{Pallot}, T.~{Colegate}, A.~{Wicenec},
  N.~{Kudryavtseva}, W.~{Arcus}, D.~{Barnes}, G.~{Bernardi}, F.~{Briggs},
  S.~{Burns}, J.~D. {Bunton}, R.~J. {Cappallo}, B.~E. {Corey}, A.~{Deshpande},
  L.~{Desouza}, B.~M. {Gaensler}, L.~J. {Greenhill}, P.~J. {Hall}, B.~J.
  {Hazelton}, D.~{Herne}, J.~N. {Hewitt}, M.~{Johnston-Hollitt}, D.~L.
  {Kaplan}, J.~C. {Kasper}, B.~B. {Kincaid}, R.~{Koenig}, E.~{Kratzenberg},
  M.~J. {Lynch}, B.~{Mckinley}, S.~R. {Mcwhirter}, E.~{Morgan}, D.~{Oberoi},
  J.~{Pathikulangara}, T.~{Prabu}, R.~A. {Remillard}, A.~E.~E. {Rogers},
  A.~{Roshi}, J.~E. {Salah}, R.~J. {Sault}, N.~{Udaya-Shankar},
  F.~{Schlagenhaufer}, K.~S. {Srivani}, J.~{Stevens}, R.~{Subrahmanyan},
  M.~{Waterson}, R.~L. {Webster}, A.~R. {Whitney}, A.~{Williams}, C.~L.
  {Williams}, and J.~S.~B. {Wyithe}, ``{The Murchison Widefield Array: The
  Square Kilometre Array Precursor at Low Radio Frequencies},'' \emph{\pasa},
  vol.~30, p.~7, Jan. 2013.

\bibitem{chime}
D.~{Castelvecchi}, ``{{\textquoteleft}Half-pipe{\textquoteright} telescope will
  probe dark energy in teen Universe},'' \emph{\nat}, vol. 523, no. 7562, pp.
  514--515, Jul 2015.

\bibitem{koo15}
L.~{Koopmans}, J.~{Pritchard}, G.~{Mellema}, J.~{Aguirre}, K.~{Ahn},
  R.~{Barkana}, I.~{van Bemmel}, G.~{Bernardi}, A.~{Bonaldi}, F.~{Briggs},
  A.~G. {de Bruyn}, T.~C. {Chang}, E.~{Chapman}, X.~{Chen}, B.~{Ciardi},
  P.~{Dayal}, A.~{Ferrara}, A.~{Fialkov}, F.~{Fiore}, K.~{Ichiki}, I.~T.
  {Illiev}, S.~{Inoue}, V.~{Jelic}, M.~{Jones}, J.~{Lazio}, U.~{Maio},
  S.~{Majumdar}, K.~J. {Mack}, A.~{Mesinger}, M.~F. {Morales}, A.~{Parsons},
  U.~L. {Pen}, M.~{Santos}, R.~{Schneider}, B.~{Semelin}, R.~S. {de Souza},
  R.~{Subrahmanyan}, T.~{Takeuchi}, H.~{Vedantham}, J.~{Wagg}, R.~{Webster},
  S.~{Wyithe}, K.~K. {Datta}, and C.~{Trott}, ``{The Cosmic Dawn and Epoch of
  Reionisation with SKA},'' \emph{Advancing Astrophysics with the Square
  Kilometre Array (AASKA14)}, p.~1, Apr. 2015.

\bibitem{mjh15}
M.~{Johnston-Hollitt}, F.~{Govoni}, R.~{Beck}, S.~{Dehghan}, L.~{Pratley},
  T.~{Akahori}, G.~{Heald}, I.~{Agudo}, A.~{Bonafede}, E.~{Carretti},
  T.~{Clarke}, S.~{Colafrancesco}, T.~A. {Ensslin}, L.~{Feretti},
  B.~{Gaensler}, M.~{Haverkorn}, S.~A. {Mao}, N.~{Oppermann}, L.~{Rudnick},
  A.~{Scaife}, D.~{Schnitzeler}, J.~{Stil}, A.~R. {Taylor}, and V.~{Vacca},
  ``{Using SKA Rotation Measures to Reveal the Mysteries of the Magnetised
  Universe},'' \emph{Advancing Astrophysics with the Square Kilometre Array
  (AASKA14)}, p.~92, 2015.

\bibitem{bea19}
A.~P. {Beardsley}, M.~{Johnston-Hollitt}, C.~M. {Trott}, J.~C. {Pober},
  J.~{Morgan}, D.~{Oberoi}, D.~L. {Kaplan}, C.~R. {Lynch}, G.~E. {Anderson},
  P.~I. {McCauley}, S.~{Croft}, C.~W. {James}, O.~I. {Wong}, C.~D. {Tremblay},
  R.~P. {Norris}, I.~H. {Cairns}, C.~J. {Lonsdale}, P.~J. {Hancock}, B.~M.
  {Gaensler}, N.~D.~R. {Bhat}, W.~{Li}, N.~{Hurley-Walker}, J.~R. {Callingham},
  N.~{Seymour}, S.~{Yoshiura}, R.~C. {Joseph}, K.~{Takahashi}, M.~{Sokolowski},
  J.~C.~A. {Miller-Jones}, J.~V. {Chauhan}, I.~{Boji{\v{c}}i{\'c}}, M.~D.
  {Filipovi{\'c}}, D.~{Leahy}, H.~{Su}, W.~W. {Tian}, S.~J. {McSweeney}, B.~W.
  {Meyers}, S.~{Kitaeff}, T.~{Vernstrom}, G.~{G{\"u}rkan}, G.~{Heald},
  M.~{Xue}, C.~J. {Riseley}, S.~W. {Duchesne}, J.~D. {Bowman}, D.~C. {Jacobs},
  B.~{Crosse}, D.~{Emrich}, T.~M.~O. {Franzen}, L.~{Horsley}, D.~{Kenney},
  M.~F. {Morales}, D.~{Pallot}, K.~{Steele}, S.~J. {Tingay}, M.~{Walker}, R.~B.
  {Wayth}, A.~{Williams}, and C.~{Wu}, ``{Science with the Murchison Widefield
  Array: Phase I results and Phase II opportunities},'' \emph{\pasa}, vol.~36,
  p. e050, Jan 2019.

\bibitem{LP18}
L.~{Pratley}, J.~D. {McEwen}, M.~{d'Avezac}, R.~E. {Carrillo}, A.~{Onose}, and
  Y.~{Wiaux}, ``{Robust sparse image reconstruction of radio interferometric
  observations with purify},'' \emph{\mnras}, vol. 473, pp. 1038--1058, Jan.
  2018.

\bibitem{PJM19a}
L.~{Pratley}, M.~{Johnston-Hollitt}, and J.~D. {McEwen}, ``{A Fast and Exact
  w-stacking and w-projection Hybrid Algorithm for Wide-field Interferometric
  Imaging},'' \emph{\apj}, vol. 874, p. 174, Apr. 2019.

\bibitem{Dabbech18}
A.~{Dabbech}, A.~{Onose}, A.~{Abdulaziz}, R.~A. {Perley}, O.~M. {Smirnov}, and
  Y.~{Wiaux}, ``{Cygnus A super-resolved via convex optimization from VLA
  data},'' \emph{\mnras}, vol. 476, pp. 2853--2866, May 2018.

\bibitem{kom15}
N.~Komodakis and J.~Pesquet, ``Playing with duality: An overview of recent
  primal?dual approaches for solving large-scale optimization problems,''
  \emph{IEEE Signal Processing Magazine}, vol.~32, no.~6, pp. 31--54, Nov 2015.

\bibitem{ono16}
A.~{Onose}, R.~E. {Carrillo}, A.~{Repetti}, J.~D. {McEwen}, J.-P. {Thiran},
  J.-C. {Pesquet}, and Y.~{Wiaux}, ``{Scalable splitting algorithms for
  big-data interferometric imaging in the SKA era},'' \emph{\mnras}, vol. 462,
  pp. 4314--4335, Nov. 2016.

\bibitem{pra19d}
L.~{Pratley}, M.~{Johnston-Hollitt}, and J.~D. {McEwen}, ``{\$w\$-stacking
  \$w\$-projection hybrid algorithm for wide-field interferometric imaging:
  implementation details and improvements},'' \emph{PASA, submitted}, Mar 2019.

\bibitem{Cornwell08}
T.~J. {Cornwell}, K.~{Golap}, and S.~{Bhatnagar}, ``{The Noncoplanar Baselines
  Effect in Radio Interferometry: The W-Projection Algorithm},'' \emph{IEEE
  Journal of Selected Topics in Signal Processing}, vol.~2, pp. 647--657, Nov.
  2008.

\bibitem{tas18}
C.~{Tasse}, B.~{Hugo}, M.~{Mirmont}, O.~{Smirnov}, M.~{Atemkeng}, L.~{Bester},
  M.~J. {Hardcastle}, R.~{Lakhoo}, S.~{Perkins}, and T.~{Shimwell}, ``{Faceting
  for direction-dependent spectral deconvolution},'' \emph{\aap}, vol. 611, p.
  A87, Apr 2018.

\bibitem{Beck13}
R.~{Beck}, J.~{Anderson}, G.~{Heald}, A.~{Horneffer}, M.~{Iacobelli},
  J.~{K{\"o}hler}, D.~{Mulcahy}, R.~{Pizzo}, A.~{Scaife}, O.~{Wucknitz}, and
  {LOFAR Magnetism Key Science Project Team}, ``{The LOFAR view of cosmic
  magnetism},'' \emph{Astronomische Nachrichten}, vol. 334, pp. 548--557, Jun.
  2013.

\bibitem{rise18}
C.~J. {Riseley}, E.~{Lenc}, C.~L. {Van Eck}, G.~{Heald}, B.~M. {Gaensler},
  C.~S. {Anderson}, P.~J. {Hancock}, N.~{Hurley-Walker}, S.~S. {Sridhar}, and
  S.~V. {White}, ``{The POlarised GLEAM Survey (POGS) I: First results from a
  low-frequency radio linear polarisation survey of the southern sky},''
  \emph{\pasa}, vol.~35, Dec. 2018.

\bibitem{bur66}
B.~J. {Burn}, ``{On the depolarization of discrete radio sources by Faraday
  dispersion},'' \emph{\mnras}, vol. 133, p.~67, Jan 1966.

\bibitem{bre05}
M.~A. {Brentjens} and A.~G. {de Bruyn}, ``{Faraday rotation measure
  synthesis},'' \emph{\aap}, vol. 441, pp. 1217--1228, Oct. 2005.

\bibitem{pratley20}
L.~{Pratley} and M.~{Johnston-Hollitt}, ``{Wide-band Rotation Measure
  Synthesis},'' \emph{arXiv e-prints}, p. arXiv:1906.00866, Jun 2019.

\bibitem{Bhat08}
S.~{Bhatnagar}, T.~J. {Cornwell}, K.~{Golap}, and J.~M. {Uson}, ``{Correcting
  direction-dependent gains in the deconvolution of radio interferometric
  images},'' \emph{\aap}, vol. 487, pp. 419--429, Aug. 2008.

\bibitem{aka16}
T.~{Akahori}, D.~{Ryu}, and B.~M. {Gaensler}, ``{Fast Radio Bursts as Probes of
  Magnetic Fields in the Intergalactic Medium},'' \emph{\apj}, vol. 824, p.
  105, Jun. 2016.

\bibitem{Shanahan19}
R.~{Shanahan}, S.~J. {Lemmer}, J.~M. {Stil}, H.~{Beuther}, Y.~{Wang},
  J.~{Soler}, L.~D. {Anderson}, F.~{Bigiel}, S.~C.~O. {Glover}, P.~{Goldsmith},
  R.~S. {Klessen}, N.~M. {McClure-Griffiths}, S.~{Reissl}, M.~{Rugel}, and
  R.~J. {Smith}, ``{Strong Excess Faraday Rotation on the Inside of the
  Sagittarius Spiral Arm},'' \emph{\apjl}, vol. 887, no.~1, p.~L7, Dec 2019.

\end{thebibliography}

\end{document}